# Direct measurement of inter-filament resistance in $Nb_3Sn$ multi-filamentary strands

L. Muzzi and V. Corato

*ENEA, Frascati Research Centre, I-00044 Frascati, Italy*

**Abstract**

In modeling the properties of superconducting multi-filamentary strands, the filament twist pitch and the inter-filament transverse resistivity play a crucial role in the definition of the current transfer length. This affects the AC losses of NbTi strands, or the transport properties of $Nb_3Sn$ wires subject to a bending strain. Although the effect of filament twist pitch on the critical current of the strand has been studied, the value of the inter-filament resistance in $Nb_3Sn$ strands is still undefined. We have performed a direct measurement of the inter-filament transverse resistance within the cross-section of a $Nb_3Sn$ multi-filamentary strand from room temperature to 4.2K. Results have been compared to the inter-filament resistance of a sample on which the outer copper stabilization layer was removed by chemical etching, obtaining interesting indications on the preferential current paths within the wire cross section.

**Introduction**

Superconducting wires used in various applications are made of very fine filaments embedded within a copper stabilization matrix[1]. The transverse inter-filament resistance is the physical parameter that governs important mechanisms occurring in the superconductor, as coupling losses in AC regimes[2] and current transfer among different filaments[3]. In the case of $Nb_3Sn$ strands, owing to the dependence of its superconducting properties on strain[4-7], the application of a mechanical bending load leads to a non-uniform distribution of $T_c$, $B_{c2}$, $J_c$ values over different filament regions within the strand cross section. For this reason, during a voltage-current measurement, mechanisms of current transfer among filaments or filament bundles occur, and the ability to redistribute the current determines the entity of the critical current degradation with bending strain. In fact, the performances of a Nb$_3$Sn strand subject to bending vary with the inter-filament transverse resistivity[8] value, which determines the current transfer length[3]. This characteristic length has to be compared with the filament twist pitch value: if the current transfer length is short with respect to the twist pitch length, which is achieved in the limit of low inter-filament resistance, then current can easily be transferred among regions of the wire cross section characterized by different superconducting properties, and the result is a moderate depletion of the overall strand critical current with bending. We have shown experimentally[9] that even an enhancement of performances can be achieved on strands subject to bending strain and axial pre-compression, if the superconducting filaments within the copper matrix are left untwisted, i.e. in the limiting case where the filament twist pitch is clearly much longer than the current transfer length (long twist pitch limit, or low inter-filament resistivity limit).

An estimation of the value of the strand transverse resistance, or at least of its order of magnitude, is not straightforward, since this parameter depends not only on the resistivity of the matrix normal metal, but also on the contact between the superconducting filaments and the matrix itself [2], as well as on the presence and layout of diffusion barriers. An experimental estimation of the transverse resistivity value can be indirectly achieved by measuring the intra-strand coupling losses in AC regimes at low frequencies[10,11], if knowing the strand twist pitch value[1,12-14]. However, the result of this measurement depends on the magnetic coupling between filaments and filament bundles, and on the resulting effective filament diameter, as well as on the wire internal structure and twist pitch[15,16].

We have performed a direct measurement of the inter-filament transverse resistivity within the cross-section of a Nb$_3$Sn multi-filamentary strand. In the present paper we report on the experiment and its main results.

**Experimental set-up**

Samples were taken from available lengths of a Nb$_3$Sn wire produced by Oxford Instruments Superconducting Technologies (OST) for ITER[17] (OST-I wire; billet #7567). The wire is characterized by a filamentary region, separated from the surrounding Cu stabilization matrix by a single tantalum (Ta) diffusion barrier. A SEM cross-section is shown in Fig. 1. The filamentary region is made of 19 filament bundles, each of which contains about 150 closely spaced superconducting filaments of about 6 μm diameter. For mechanical support, the wires have been inserted into resin and cut in 3.5 mm long pieces. After cutting, the samples have been carefully treated by a diamond lapping paste on both sides, in order to avoid effects of filament coalescence and fictitious superconducting short circuits. Electrical contacts are attached on the sample surface by micro-bonding of thin Al wires, as shown in Fig. 2a (sample $\alpha$). Each current contact is coupled to a voltage contact placed in a different position over the same filament bundle, in order to avoid measuring the contribution of the bonding-to-superconductor contact resistance. Considering the level of microscopic resolution that can be achieved with the present technique, by which electrical contact extend over several filaments, each of the 19 filament bundles is treated as a single superconducting macro-filament. Each Al bonding wire is connected to a golden track limited by a terminal pad, where the proper measurement wire is welded.

On one sample an acid etching has been carried out, aimed at removing only the outer matrix copper sheath, as shown in Fig. 2b (sample $\beta$). This was done in order to prove to what extent this surrounding layer represents a preferential path, for current transfer among filament bundles.

The test strand is mounted on a copper sample holder, to which a Cernox thermometer is attached in order to acquire the temperature dependence. The measurements have been performed in a He gas flow cryostat, with the sample shielded against the direct exposure to the cooling He gas flow by means of a copper box. A four-point technique has been applied for the resistance measurement, by means of a Keithley high precision current supply, coupled to a Keithley nanoVoltmeter for the voltage signal, with appropriate filters activated. At each temperature, current is automatically ramped at steps, in the range -500 *mA* up to 500 *mA*, and the voltage is acquired at each step. The resistance is extracted from the angular coefficient of the obtained V-I curve. Fig. 3 shows an example of experimental V-I curves measured between the voltage pair A-C in the sample $\alpha$ (see Fig. 2a) from 4K to 9K. The determination of the resistance is easily obtained by a linear fit of the data at 9K, providing a value of about 220 n$\Omega$, while at lower temperatures it's more problematic, due to the high non-linearity of the V-I characteristics. Nevertheless, considering the asymptotic behavior of the tails at 4.2K and 6K toward the curve measured at 9K, we assume that, below this temperature, the effective transverse resistance has the same value as that determined at 9K. As a matter of fact, we believe that the non-linear behavior of the V-I curve, typical of the superconductors, can be attributed to a proximity effect coupling[18,19] within the wire filamentary region. An effect due to the presence of possible un-reacted niobium regions and/or of the Ta diffusion barrier, that becomes superconducting below 4.5K, could play a role as well. Non-linear characteristics have also been evidenced at temperatures close to the Nb$_3$Sn critical temperature (around 18K): in this case we have extracted the resistance from the superconducting branch, before the Nb$_3$Sn transition occurs. Finally, for higher temperatures the characteristics are straight lines, because of the ohmic behavior, and the definition of the resistance value is straightforward.

**Measurement results**

Fig. 4 reports on the transverse resistance versus temperature curve, measured on the OST-I wire (sample $\alpha$), between bundles A-B (stars) and A-C (circles). The very low value of the resistance ratio between 300K and 25K observed on the transverse resistance is to be compared with a value of 156, previously measured along the strand[20], as mainly due to the stabilization matrix normal conductance. This gives a hint that a relevant contribution to the transverse resistance comes from the contact resistance between superconducting filaments and normal matrix, more than by the Cu resistivity itself. However, open triangles in Fig. 4 show that a clear resistance increase is observed if the external pure Cu stabilization annulus is removed by chemical etching (sample β), underlying that this outer region is certainly a preferential path for the current exchange among filament bundles. This was indirectly observed by other authors[21-25], who advocated this effect in order to obtain a satisfactory agreement between simulations and measurements of critical currents as function of bending strain.

In addition, Fig. 4 shows that a drop of about 2÷3 orders of magnitude in resistance is observed around 18K, corresponding to the Nb$_3$Sn superconducting transition temperature. This demonstrates that the current path within the wire cross-section changes completely when the filaments become superconducting and partly flows through the superconductor itself. We hypothesize that a percolative current path through filaments within a bundle and across next bundle establishes, due to a proximity coupling effect between closely spaced filaments[18,19]. It is our intention to probe this hypothesis by accurate measurements of the V-I curves shape in the range 4.2K ÷ 15K in the presence of a background magnetic field.

**Discussion**

A description of the inter-filament resistivity in technological superconducting strands has been formulated by Wilson[1], based on the computation of the circulating currents excited within the wire by AC varying fields. In his treating, the effective transverse resistivity of the matrix $\rho_{et}$ can be described by the parallel contribution of the resistivity encountered by currents flowing partially through superconducting paths, either through the filamentary region ($\rho_f$) or in the peripheral shells ($\rho_p$), and of the resistivity encountered by normal eddy currents flowing in the outer layers ($\rho_{eddy}^{Cu}$ and $\rho_{eddy}^{Ta}$ for copper and tantalum sheaths, respectively)[26]:

$$\frac{1}{\rho_{et}} = \frac{1}{\rho_f} + \frac{1}{\rho_p} + \frac{1}{\rho_{eddy}^{Cu}} + \frac{1}{\rho_{eddy}^{Ta}} \qquad (1)$$

The resistivity experienced by the eddy currents and that due to superconducting currents flowing partially through peripheral layers, are given by the following relations:

$$\frac{1}{\rho_{eddy}} = \frac{a_f w}{\rho}\left(\frac{2\pi}{L}\right)^2 \qquad (2)$$

$$\frac{1}{\rho_p} = \frac{w}{a_f \rho} \qquad (3)$$

where $a_f = 0.275$ mm is the radius of the filamentary region, $w$ is the width of the layer of resistivity $\rho$ and $L$ the filament twist pitch that, in our case, corresponds to the sample height $h = 3.5$ mm.

The last term in Eq. (1), connected to the presence of the Ta diffusion barrier, can be neglected with respect to the contribution of copper, $1/\rho_{eddy}^{Cu}$, because of the higher resistivity and of the smaller width of the tantalum layer, as compared to the Cu outer shell.

It should be considered that in the measurement of the etched strand (sample $\beta$) the outer copper layer is absent, so that $\rho_\beta^{exp}$ can be written as[1]:

$$\frac{1}{\rho_\beta^{exp}} = \frac{1}{\rho_f} + \frac{w_1}{a_f \rho_{Ta}} + \frac{a_f w_1}{\rho_{Ta}}\left(\frac{2\pi}{h}\right)^2 \qquad (4)$$

where $w_1 = a_1 - a_f = 10$ μm is the tantalum barrier width, having a resistivity $\rho_{Ta} = 3.6 \cdot 10^{-7}\,\Omega \cdot m$ at 4.2K. Considering that at cryogenic temperatures we measured $\frac{1}{\rho_\beta^{exp}} \approx 2.4 \cdot 10^8\,(\Omega \cdot m)^{-1}$, while the sum of the second and third terms on the right side of Eq. (4) can be computed, giving about $1.3 \cdot 10^5\,(\Omega \cdot m)^{-1}$, it's clear that the most important in (4) contribution is due to transverse resistivity in the filamentary zone, so that $\rho_f$ can be approximated by the transverse resistivity measured on sample $\beta$: $\rho_f \approx \rho_\beta^{exp}$.

A crucial role is played by the term $\rho_p$ in (1), which should take into account the two concentric layers surrounding the filamentary zone: the thin tantalum barrier and the copper outer shell. As reported in [27], the resistivity in this particular case can be expressed by the following relation:

$$\frac{1}{\rho_p} = \frac{1}{\rho_{Ta}}\left(\frac{a_f}{w_1}\right) + \frac{1}{\rho_{Cu}}\left(\frac{\rho_{Cu}}{\rho_{Ta}}\right)^2 \frac{a_1 a_2^2}{w_1^2 w_2} \tag{5}$$

Where $a_1 = 0.285$ mm and $a_2 = 0.405$ mm represent the radii of the region included in the Ta barrier and of the whole strand, respectively (see Fig. 1), and $w_2 = a_2 - a_1 = 0.120$ mm is the width of the copper layer, having a resistivity $\rho_{Cu} \approx 1.0 \cdot 10^{-10}\ \Omega \cdot m$ for RRR=156 at 4.2K.

Summarizing, the effective transverse resistivity of the matrix is reported below:

$$\frac{1}{\rho_{et}} = \frac{1}{\rho_f} + \frac{1}{\rho_{Ta}}\left(\frac{a_f}{w_1}\right) + \frac{1}{\rho_{Cu}}\left(\frac{\rho_{Cu}}{\rho_{Ta}}\right)^2 \frac{a_1 a_2^2}{w_1^2 w_2} + \frac{a_f w_2}{\rho_{Cu}}\left(\frac{2\pi}{h}\right)^2 \tag{6}$$

In our experiment, we have directly measured $\rho_{et}$ on sample $\alpha$, and the measurement at 4.2K of the sample without the peripheral copper layer gives $\rho_\beta^{exp} \approx \rho_f = 4.2 \cdot 10^{-9}\ \Omega \cdot m$. Making use of Eq. (6), with the parameters reported above, we can thus estimate the expected value for the effective transverse resistivity $\rho_{et} = 7.3 \cdot 10^{-10}\ \Omega \cdot m$, that is in perfect agreement with the experimental one $\rho_\alpha^{exp} = \rho_{et} = 7.5 \cdot 10^{-10}\ \Omega \cdot m$.

As one can see, owing to the high contact resistance at the filament-to-matrix interface, caused by inter-metallic layers formed during the Nb$_3$Sn reaction heat treatment, the transverse resistivity across the filamentary region is much larger than $\rho_{Cu}$. This might be due to the fact that, after the heat treatment, tin has diffused throughout the copper matrix within the filamentary region, while the outer stabilization region is protected by the tantalum diffusion barrier. However, this same effect has been also observed in NbTi wires[28], where no reaction heat treatment is performed.

Carr[10] has proposed a formula to describe the increase of the transverse resistivity across the filamentary region due to the presence of high contact resistance:

$$\rho_f = \rho_{Cu} \frac{(1+\lambda)}{(1-\lambda)} \tag{7}$$

Where $\lambda$ is the fraction of superconductor in the cross-section. For the OST-I strand $\lambda$ = 0.5, then the transversal resistivity should be about 3 times larger than $\rho_{Cu}$, while the experimental measurement gives a factor 42. This phenomenon can be partially explained if the size effect of copper resistivity[2] is taken into account. In fact, in the OST-I wire, the filaments (of about 6 μm thickness) are separated of about 1 μm, as shown in Fig. 5, that reports a SEM image of a sample, the surface of which has been etched with acid in order to remove the matrix material. This figures also confirms that the acid etching keeps the Ta barrier unchanged. If $\rho_{Cu}$ in Eq. (7) is substituted with the computed copper resistivity in the presence of the size effect, computed for a Cu RRR=156, $\rho_{Cu}^{size} = 7.4 \cdot 10^{-10}\, \Omega \cdot m$, the calculated transversal resistivity in the filamentary region at 4.2K turns out to be $2.2 \cdot 10^{-9}\, \Omega \cdot m$ that is lower, even if in the same order of magnitude, than the measured value $\rho_\beta^{exp} \approx \rho_f = 4.2 \cdot 10^{-9}\, \Omega \cdot m$. The remaining difference can finally be justified considering that during the reaction heat treatment, within the filamentary region tin diffuses into the Cu matrix, which clearly leads to an increase of its resistivity.

From the above discussions, we infer that both the size effect of the inter-filament Cu resistivity and a high contact resistance between filaments and matrix substantially contribute to the $\rho_{Cu}$ value.

**Conclusions and perspectives**

We have performed a direct measurement of the inter-filament resistance in Nb$_3$Sn multi-filamentary wires. Compared to the resistivity of the stabilization copper, a large value of transversal resistivity has been measured at zero applied magnetic field in the filamentary region of a sample wire. The phenomenon can be reasonably explained by considering a high contact resistance between filaments and matrix, as well as the size effect of the Cu resistivity. Furthermore, different mechanisms of current transfer have been evidenced, including a relevant percolative current path through the superconducting filaments, due to a proximity coupling effect, which disappears at the $Nb_3Sn$ transition to normal state.

These preliminary measurements of the effective transverse resistivity proved very promising, and thus further developments are foreseen to optimize the experimental set-up. In particular, some modification are foreseen, in order to allow similar characterizations also in the presence of a background magnetic field. This will: a) cause a magnetic coupling between neighboring filaments; b) help in clarifying the possible role of unreacted Nb regions and of the Ta diffusion barrier; c) help in clarifying the balance between contact resistance contributions and contributions due to proper Cu (or bronze) resistances; d) allow to clarify the effect of proximity coupling.

This measurement procedure will be extended to other $Nb_3Sn$ strands, with different layout of filaments, diffusion barriers, production techniques (powder-in-tube or bronze route) as well as to NbTi wires, in order to clarify further aspects of current transfer processes within wire cross-sections.

**Acknowledgment**

The authors acknowledge the contribution of A. Rufoloni for SEM images and A. Augieri for the measurement set-up and data acquisition system.

**References**

[1] M. N. Wilson, Superconducting magnets, Oxford University Press (1983).

[2] M. N.Wilson, Cryogenics **48**, 381 (2008).

[3] J. W. Ekin, J. Appl. Phys. **49,** 3406 (1978); J. W. Ekin, J. Appl. Phys. **49,** 3410 (1978).

[4] R. Flukiger, W. Schauer, W. Specking, B. Schmidt and E. Springer, IEEE Trans. Magn. **17,** 2285 (1981).

[5] C. L. Snead, and M. Suenaga, Appl. Phys. Lett. **37,** 659 (1980).

[6] D. O. Welch, Adv. Cryog. Eng. (Materials) **26,** 48 (1980).

[7] A. Godeke, Supercond. Sci. Technol. **19,** R68 (2006).

[8] J. W. Ekin "Strain scaling law and the prediction of uniaxial and bending strain effects in multifilamentary superconductors" in Filamentary A15 Superconductors, M. Suenaga, AF Clark, Ed. Plenum Press (1980).

[9] L. Muzzi, V. Corato, R. Viola and A. della Corte, J. Appl. Phys. **103,** 073915 (2008).

[10] W. Carr, IEEE Trans. Magn. **13,** 192 (1977).

[11] G. Ries, IEEE Trans. Magn. **13**, 524 (1977).

[12] A. Nijhuis, N. C. van den Eijnden, Y. Ilyin, E. G. van Putten, G. J. T. Veening, W. A. J. Wessel, A. den Ouden and H. H. J. ten Kate, Supercond. Sci. Technol. **18,** S273 (2005).

[13] A. Nijhuis, Y. Ilyin, W. Abbas and W. A. J. Wessel, IEEE Trans. on Appl. Supercond. **17**, 2680 (2007).

[14] M. D. Sumption, D. S. Pyun and E. W. Collings, IEEE Trans. on Appl. Supercond. **3,** 859 (1993).

[15] M. D. Sumption, E. Lee, S. X. Dou and E. W. Collings, Physica C **335**, 164 (2000).

[16] R. B. Goldfarb and K. Itoh, J. Appl. Phys. **75,** 2115 (1994).

[17] J. A. Parrell, M. B. Field, Y. Zhang and S. Hong IEEE Trans. on Appl. Supercond. **15,** 1200 (2005).

[18] M. D. Sumption and E. W. Collings Cryogenics **34,** 491 (1994).

[19] K. Yasohama, S. Nagano, Y. Kubota and T. Ogasawara IEEE Trans. on Appl. Supercond. **5,** 729 (1995).

[20] L. Muzzi, S. Chiarelli, A. della Corte, A. Di Zenobio, M. Moroni, A. Rufoloni, A. Vannozzi, E. Salpietro and A. Vostner, IEEE Trans. on Appl. Supercond. **16,** 1253 (2006).

[21] C. Fiamozzi Zignani, V. Corato, A. della Corte, A. Di Zenobio, G. Messina and L. Muzzi, IEEE Trans. on Appl. Supercond., in press (2008).

[22] H. Murakami, A. Ishiyama, H. Ueda, N. Koizumi and K. Okuno, IEEE Trans. on Appl. Supercond. **17,** 1394 (2006).

[23] P. L. Ribani, D. P. Boso, M. Lefik, Y. Nunoya, L. Savoldi Richard, B. A. Schrefler and R. Zanino, IEEE Trans. on Appl. Supercond. **16,** 860 (2006).

[24] M. Hirohashi, H. Murakami, A. Ishiyama, H. Ueda, N. Koizumi and K. Okuno, IEEE Trans. Appl. Supercond. **16,** 1721 (2006).

[25] R. Zanino, D. P. Boso, M. Lefik, P. L. Ribani, L. Savoldi Richard and B. A. Schrefler, IEEE Trans. Appl. Supercond. **18,** 1067 (2008).

[26] Regarding the "eddy currents" contributions in (1), we kept the same nomenclature as in [1] for coherence, even if in our case no eddy currents, as excited by AC fields, properly exist.

[27] B. Turck, J. Appl. Phys. **50**, 5397 (1979).

[28] B. Turck, M. Wake, M. Kobayashi, Cryogenics **17**, 217 (1977).

**Figures**

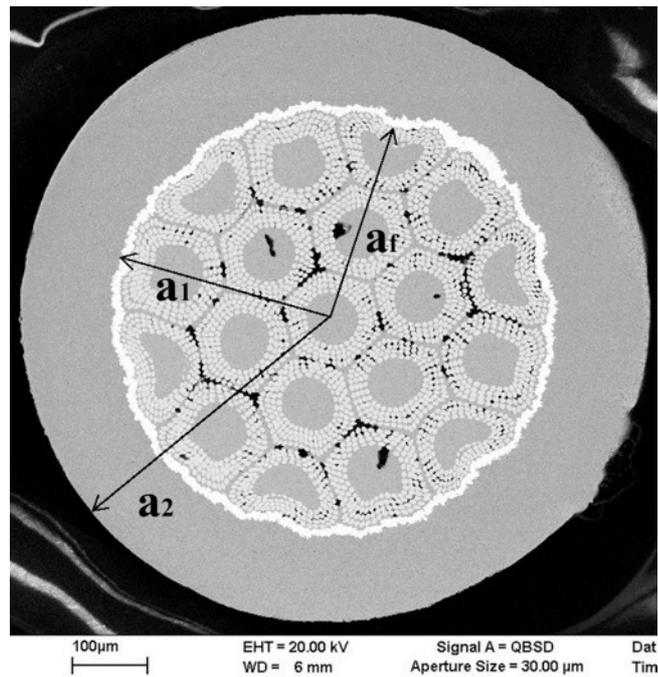

Fig. 1. SEM image of the cross section of the OST-I strand. The radii of the filamentary region $a_f$, of the zone included in the tantalum barrier ($a_1$) and of the whole strand ($a_2$) are indicated.

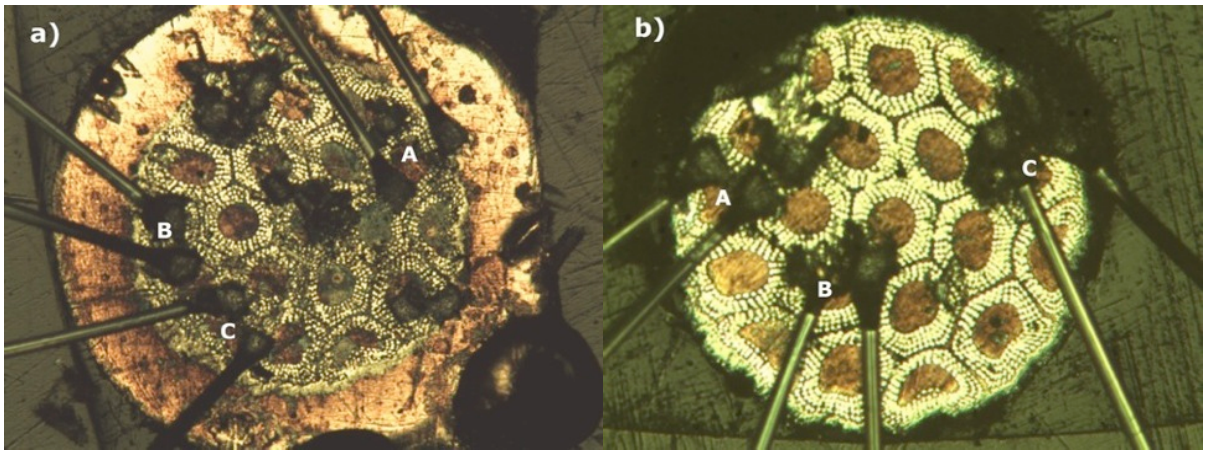

Fig. 2. Photograph of a) OST-I strand (sample $\alpha$) with electrical contacts attached on the sample surface by micro-bonding of thin Al wires; b) the same strand with the outer matrix copper sheath removed by chemical etching (sample $\beta$).

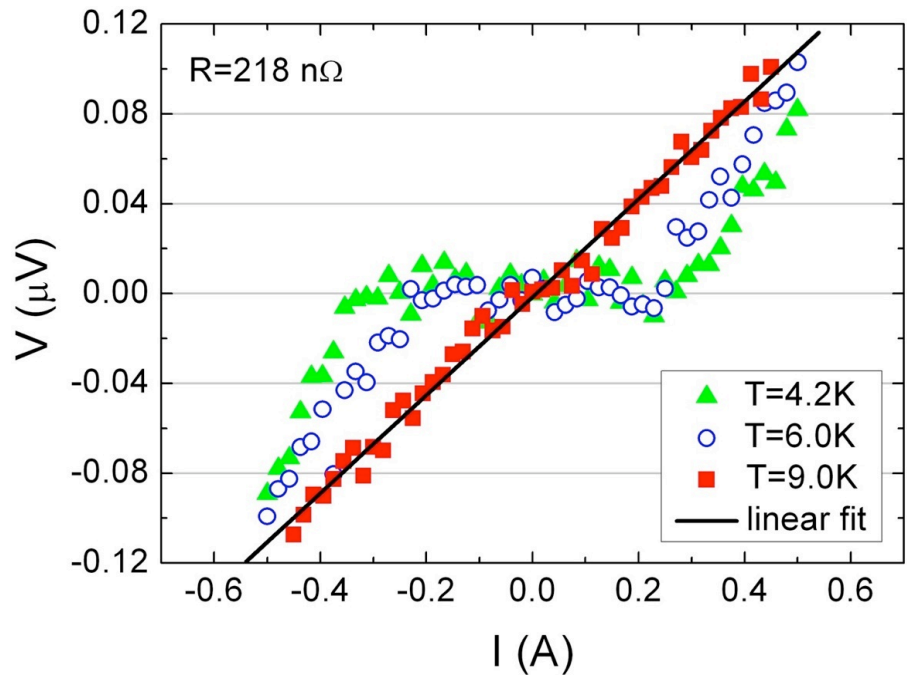

Fig. 3. Experimental V-I characteristics measured between the bundles A-C on sample α, at 4.2K (triangles), 6K (open circles) and 9K (squares).

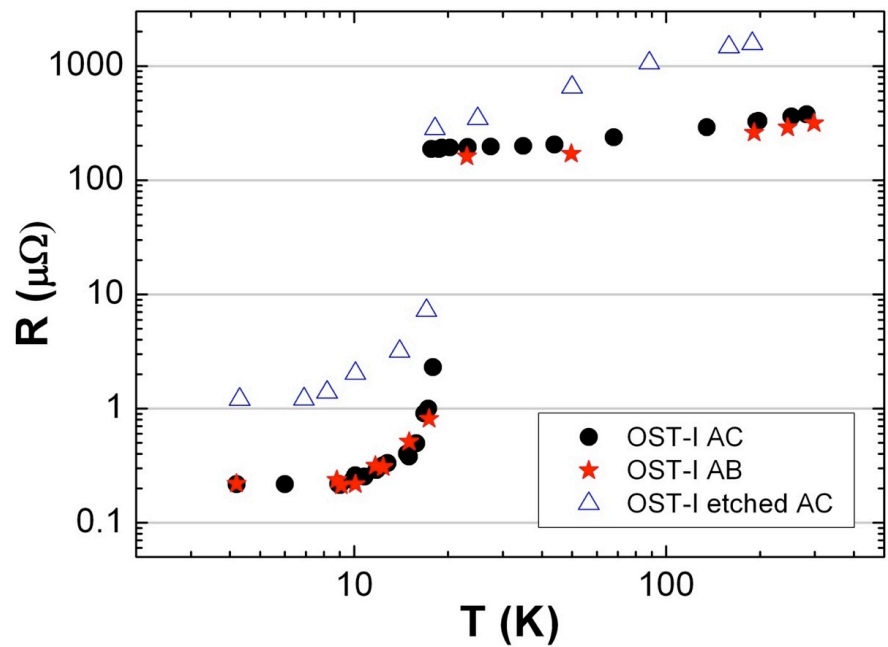

Fig. 4. Transverse resistance vs. temperature curve, measured on the OST-I wire (sample α), between bundles A-B (stars) and A-C (circles). Open triangles refer to the resistance measured on sample β between the voltage tap pair A-C. The resistance increase, observed when the external pure Cu layer is removed, evidences that this outer region is a fundamental path for the current exchange among filament bundles.

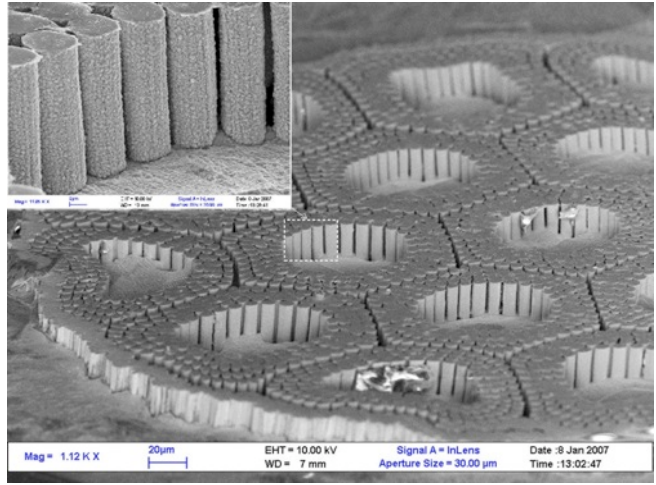

Fig. 5. SEM image of the sample, whose surface has been etched with acid in order to remove the matrix material. The single filaments (of about 6 μm thickness), having a distance of about 1 μm, are visible in the inset.